\documentclass[
jpsj,
twocolumn,
showpacs,
showkeys
]{revtex4-1}
\usepackage{amsthm}
\theoremstyle{plain}
\newtheorem{theorem}{Theorem}
\newtheorem*{theorem*}{Theorem}

\newtheorem*{definition*}{Definition}

\newtheorem*{lemma*}{Lemma}
\newtheorem{corollary}[theorem]{Corollary}

\usepackage{physics}

\usepackage{braket}
\usepackage{delarray}
\usepackage{here}
\usepackage{txfonts}

\usepackage[dvipdfmx]{graphicx}
\usepackage{bm}
\usepackage{color}
\usepackage{ulem}

\newcommand{\be}{\begin{eqnarray}}
\newcommand{\ee}{\end{eqnarray}}
\newcommand{\ba}{\begin{array}}
\newcommand{\ea}{\end{array}}
\newcommand{\bmat}{\left(\begin{array}}
\newcommand{\emat}{\end{array}\right)}
\newcommand{\no}{\nonumber}

\begin{document}
\title{Threshold theorem in quantum annealing with deterministic analog control errors}
\author{Manaka Okuyama$^1$}
\author{Masayuki Ohzeki$^{1,2,3}$}
\affiliation{$^1$Graduate School of Information Sciences, Tohoku University, Sendai 980-8579, Japan}
\affiliation{$^2$Department of Physics, Tokyo Institute of Technology, Oh-okayama, Meguro-ku, Tokyo,152-8551, Japan}
\affiliation{$^3$Sigma-i Co., Ltd., Tokyo 108-0075, Japan} 

\begin{abstract} 
We investigate the effect of deterministic analog control errors in the time-dependent Hamiltonian on isolated quantum dynamics.
Deterministic analog control errors are formulated as time-dependent operators in the Schr\"odinger equation.
We give an upper bound on the distance between two states in time evolution with and without deterministic analog control errors.
As a result, we prove that, if the strength of deterministic analog control errors is less than the inverse of computational time, the final state in quantum dynamics without deterministic analog control errors can be obtained through a constant-order number of measurements in quantum dynamics with deterministic analog control errors.

\end{abstract}
\date{\today}
\maketitle

\section{Introduction}
Quantum annealing~\cite{KN,RCC,BBRA,BRA,SMTC,FGGS,ADKLLR,MLM} is an analog quantum computation that utilizes continuous time evolution of quantum systems, and, thereby, analog control errors of the parameters are inevitable in experimental systems.
Because the theory of quantum error correction and suppression is incomplete in quantum annealing~\cite{JFS,Lidar,QL,PAL,VMKOBS}, estimating the effect of analog control errors is one of the most critical problems.

There are two main types of analog control errors in quantum annealing.
One is a stochastic control error~\cite{WZ,GZ,DRC,ACHQGHLG}, which represents an instantaneous parameter fluctuation.
For this type of control error, recent studies~\cite{OOO,Kobayashi} proved that, if the strength of the stochastic control errors is less than the inverse of the computation time, information about the final state in quantum dynamics without analog control errors can be recovered from quantum dynamics with stochastic control errors.
The other is deterministic control error, which is, for example, a bias acting on the magnetic field or a deviation in the value of the interaction.
Deterministic control errors have been discussed so far in many literatures~\cite{YKL,CFP,MAL,MGA,RC,PHL,AMH}, but they are limited to specific problems.

The present study investigates in general whether it is possible to recover information about the target state, which is the final state in ideal time evolution, from quantum dynamics with deterministic analog control errors. 
We give an upper bound on the distance between two states in quantum dynamics with and without deterministic control errors using only information about the deterministic control errors.
Furthermore, using this bound, we prove that, if the strength of the deterministic control errors is less than the inverse of the computation time, information about the target state  can be recovered through a constant-order number of measurements in quantum dynamics with deterministic analog control errors.
This result is intuitively obvious but it is important from the perspective of experimental systems to give mathematical proof.
The proof is based on the method proposed by Kieu to derive a quantum speed limit~\cite{Kieu,OO}.

The organization of this paper is as follows.
In Sec. II, we define the model and obtain the main result. 
Finally, our conclusion is presented in Sec. III.

\section{Result}
We consider the following isolated quantum dynamics:
\be
i \frac{d}{dt} |\psi(t)\rangle&=& \hat{H}(t) |\psi(t)\rangle, \label{error-less-Sch}
\ee
where $0\le t\le T$ and $\hbar=1$.
In general, it is difficult to completely control the time-dependent Hamiltonian $\hat{H}(t)$ without control errors in experimental systems.
Deterministic analog control errors can take any form physically permissible but should also be described as a Hermitian operator since we consider isolated quantum dynamics.
Thus, we incorporate the deterministic analog control errors of $\hat{H}(t)$ into the Schr\"odinger equation as a Hermitian operator $\hat{V}(t)$.
We express the Schr\"odinger equation with deterministic analog control errors as follows:
\be
i \frac{d}{dt} |\phi(t)\rangle&=&( \hat{H}(t)+\hat{V}(t)) |\phi(t)\rangle.\label{error-Sch}
\ee
Then, we obtain the following result.
\begin{theorem} \label{th2}
The distance between two final states $\ket{\psi(T)}$ and $\ket{\phi(T)}$ is bounded from above by
\be
\| \ket{\psi(T)}- \ket{\phi(T)} \|  \le  v  , \label{main-result}
\ee
where $\| \ket{a}\|\equiv\sqrt{\braket{a|a}}$, $v\equiv \int_0^T dt \left\| \hat{V}(t)  \right\|$, and $\left\| \hat{A}  \right\|$ is the eigenvalue of $\hat{A}$ with the largest absolute value.
\end{theorem} 
\begin{proof}[Proof of Theorem 1]
From Eqs. (\ref{error-less-Sch}) and (\ref{error-Sch}), we obtain
\be
 \dv{}{t} \qty(\ket{\psi(t)}-\ket{\phi(t)})&=& -i \hat{H}(t)\qty(\ket{\psi(t)}-\ket{\phi(t)}) + i\hat{V}(t) \ket{\phi(t)},
\ee
and
\be
\dv{}{t}\| \ket{\psi(t)}- \ket{\phi(t)} \|^2
&=&2\Re \qty{ (\bra{\psi(t)}- \bra{\phi(t)} ) \dv{}{t}\qty(\ket{\psi(t)}- \ket{\phi(t)})  }
\no\\
&=&2\Re \qty{ (\bra{\psi(t)}- \bra{\phi(t)} )  i\hat{V}(t) \ket{\phi(t)}  }
\no\\
&\le&2 \| \ket{\psi(t)}- \ket{\phi(t)} \| \cdot\| \hat{V}(t) \ket{\phi(t)} \|,
\ee
where we used the Cauchy-Schwartz inequality.
On the other hand, we find
\be
\dv{}{t}\| \ket{\psi(t)}- \ket{\phi(t)} \|^2 &=&2\| \ket{\psi(t)}- \ket{\phi(t)} \| \cdot \dv{}{t}\| \ket{\psi(t)}- \ket{\phi(t)} \| .\no\\
\ee
Thus, we obtain
\be
\dv{}{t}\| \ket{\psi(t)}- \ket{\phi(t)} \| &\le& \| \hat{V}(t) \ket{\phi(t)} \| 
\le  \| \hat{V}(t)  \|.
\ee
Finally, by integrating both sides from $0$ to $T$, we arrive at Eq. (\ref{main-result}).
\end{proof}
It is worth mentioning that the right hand side of Eq. (\ref{main-result}) contains only information about the control errors $\hat{V}$ and not about $\hat{H}(t)$.

The inequality (\ref{main-result}) makes sense only if $v< 2$ is satisfied because 
\be
\| \ket{\psi(T)}- \ket{\phi(T)} \|=\sqrt{2-2\Re \braket{\psi(t)|\phi(t)} } \le  2.
\ee
In particular, when the strength of deterministic control errors is less than the inverse of the computation time, 
\be
\left\| \hat{V}(t)  \right\| < \frac{\sqrt{2}}{T},
\ee
 we have
\be
\| \ket{\psi(T)}- \ket{\phi(T)} \| \le v   <\sqrt{2} .
\ee
This means that the two final states have non-zero overlap
\be
\Re \braket{\psi(T)|\phi(T)}\ge 1-\frac{v^2}{2} >0. \label{overlap-bound}
\ee
Then, it is possible to recover the information about $\ket{\psi(T)}$ from $\ket{\phi(T)}$.

For example, we expand the two final states $\ket{\psi(T)}$ and $\ket{\phi(T)}$ as
\be
\ket{\psi(T)}=\sum_n C_n\ket{n},
\\
\ket{\phi(T)}=\sum_n D_n\ket{n},
\ee
where $\ket{n}$ is the measurement basis. 
We are interested in the $m$th eigenstate of the measurement basis and its probability amplitude $C_m$ is given by
\be
|C_m|^2=1-\epsilon^2,
\ee
with $0\le\epsilon <1$.
Then, we arrive at:
\begin{corollary}
If 
\be
1-v^2/2 >\epsilon\ge0, \label{v-e-cond}
\ee
then the probability amplitude of the $m$th eigenstate in the Schr\"odinger equation with deterministic analog control errors (\ref{error-Sch}) takes a non-zero value,
\be
|D_m|\ge\frac{1-\frac{v^2}{2}-\epsilon}{\sqrt{1-\epsilon^2}} >0.
\ee
\end{corollary}
Corollary 2 states that the number of measurements required to obtain $\ket{m}$ is independent of the computation time $T$ in quantum dynamics with deterministic analog control errors (\ref{error-Sch}).
Thus, under the condition (\ref{v-e-cond}), deterministic control errors do not seriously affect the efficiency of quantum annealing.

The condition (\ref{v-e-cond}) can be rewritten as
\be
\int_0^T dt \left\| \hat{V}(t)  \right\| < \sqrt{2(1-\epsilon)}.
\ee
It may seem difficult to satisfy this condition for large $T$.
However, when $T$ is large, the parameters should change slowly and the strength of the analog control errors is expected to be smaller.
Thus, the condition (\ref{v-e-cond}) is not far from experimental systems and may be acceptable.
\begin{proof}[Proof of Corollary 2]
From Eq. (\ref{overlap-bound}), we obtain
\be 
0&<&1-\frac{v^2}{2} \le \Re \braket{\psi(T)|\phi(T)} 
\le |\braket{\psi(T)|\phi(T)}|
\no\\
&\le& \sum_{n} |C_n D_n|
= \sqrt{1-\epsilon^2} |D_m|+\sum_{n(\neq m)} |C_n D_n|
\no\\
&\le& \sqrt{1-\epsilon^2} |D_m|+\epsilon \sqrt{1-|D_m|^2}| 
\no\\
&\le& \sqrt{1-\epsilon^2} |D_m|+\epsilon,
\ee
where we used the Cauchy-Schwartz inequality.
Thus, using Eq. (\ref{v-e-cond}), we obtain
\be
|D_m|\ge\frac{1-\frac{v^2}{2}-\epsilon}{\sqrt{1-\epsilon^2}} >0.
\ee
\end{proof}
\section{Conclusions}

We have established a threshold theorem that provides a sufficient condition for obtaining the target state in isolated quantum dynamics with any deterministic analog control error.

We have considered only deterministic analog control errors.
A similar threshold theorem for stochastic analog control errors has already been obtained in Ref. \cite{OOO}.
For both types of analog control error, the same point is that, if the strength of the control errors is less than the inverse of the computation time, the target state can be obtained through a constant-order number of measurements in quantum dynamics with analog control errors.
It is an interesting future problem to combine these results.

Finally, we emphasize that we do not impose any assumptions on time evolution.
Considering a specific schedule for each problem, such as adiabatic time evolution, might improve the present results.


The present work was financially supported by JSPS KAKENHI Grant No. 19H01095,  20H02168 and 21K13848.


\end{document}